\documentclass{article}

\usepackage{amsmath}
\usepackage{graphicx}

\begin{document}

\title{Replicator dynamics with mutations for games with a continuous
strategy space}
\author{M.Ruijgrok$\thanks{%
Mathematics Institute, Utrecht University, Utrecht, The Netherlands. e-mail
address: ruijgrok@math.uu.nl}$ and Th.W. Ruijgrok$\thanks{%
Institute of Theoretical Physics, Utrecht University, Utrecht, The
Netherlands.}$}
\date{May 2005}
\maketitle

\begin{abstract}
A partial differential equation is derived, describing the replicator
dynamics with mutations of games with a continuous strategy space. This
equation is then applied to continuous versions of symmetric 2x2 games, such
as the Prisoners Dilemma, Hawk-Dove and Coordination games, and to the
Ultimatum Game. In the latter case, we find that adding even a small mutation
term to the replicator equation leads to a solution where the average offer is
significantly larger than zero.
\end{abstract}

\section{Introduction}

Evolutionary game dynamics is a fast developing field, with applications in
biology, economics, sociology and anthropology. Background material and
countless references can be found in the monographs by Weibull \cite{WB},
Fudenburg \& Levine \cite{FL}, Samuelson \cite{SA}, Hofbauer \& Sigmund \cite%
{HS1}, Gintis \cite {GIN}, Cressman \cite{CRS} and Vincent \& Brown \cite{VB} or in the survey
paper by Hofbauer \& Sigmund \cite{HS2}. The standard ingredients of
evolutionary game dynamics are a population of players, an $n$-person game, a
set of strategies and a rule to update the distribution of strategies from
one generation to the next. Within this general setting several variations
are possible: time can be discrete or continuous, populations can be finite
or infinite, the game can have a finite or infinite number of strategies.
Also, there are several choices for the updating rule, the most popular of
which are Adaptive Dynamics (See e.g. Diekmann \cite{DI}) and Replicator
Dynamics, introduced by Taylor \& Jonker \cite{TJ}. In this paper we
consider a model with continuous time, an infinite population and a 2-person
game, where each participant has a continuum of strategies to choose from.
The update rule we use is the replicator dynamics, with deterministic
mutations. Although this model has been alluded to by Dieckmann \cite{DIC},
where a hierarchy of evolutionary models is presented, its details have not
been worked out before.

The strategy space that we consider is a subset of $R^{n}$, which in all
applications is compact. In section 2 we first consider replicator dynamics
with a deterministic mutation term on a finite set of strategies. Using a
procedure that is familiar in Statistical Mechanics, we make a transition to
an infinite, continuous, strategy-space. The state-space of the model is
then no longer discrete, but consists of distributions over the strategies.
We derive a partial differential equation, together with appropriate
boundary conditions, that describes the time-evolution of this distribution,
given an initial distribution. Without the mutation term, the equation we
derive was previously studied by Cressman \cite{CR} and Oechsler \& Riedel 
\cite{OR}. Our addition of a mutation term is new and leads to completely
different dynamics.

In section 3, the equation is applied to symmetric 2x2 games, where we need
to extend the original two-strategy set to a continuous one. This can be
done by allowing for strategies that play one of the pure strategies with a
certain probability. We compare our results with those of \ Vaughan \cite{VA}%
, who analyses the replicator dynamics for the pure strategies, to which a
stochastic perturbation term is added, leading to a Fokker-Planck equation.
There are similarities between the two models, but also striking
differences. In the stochastic mutation version, there is always only one
attracting, stationary, distribution, which for small mutations converges to
a point-distribution (Dirac-delta function). In other words, all players
eventually use the same strategy. In the deterministic mutation version,
however, we find the possibility of two attracting stationary distributions
existing simultaneously. Also, in certain cases the limiting distribution
for small mutations is not concentrated on a point, but has the whole
strategy space as support.

A more complicated example, namely the Ultimatum Game, is treated in section
4. In this game, a sum of money is to be split by two players. The first
player proposes the split and the second player then has the choice to
accept the split or refuse, in which case both players get nothing. The
solution offered by standard game theory is for the second player to accept
any amount (something is always better than nothing) and for the first
player, therefore, to offer the lowest possible amount. In our model we find
that the offers converge to a Gauss-like distribution around a mean that is
not equal to zero. The position of the mean and the width of the
distribution seem to converge to zero when the rate of mutation vanishes.
However, this convergenge is slow, so that even for very small values of the
mutation rate, the average offer is well above zero. Also, the dynamics
shows two time-scales. A random starting distribution is initially attracted
to a member of a certain set of distributions, which can have an average
offer much larger than zero. Then, on a time-scale inversely proportional to
the rate of mutation, the solution converges to the, unique, stationary
solution.

In section 5 we discuss the results and suggest some topics for further
research.

\section{From the discrete to the continuous equation}

The equation for the replicator dynamics with continuous strategy space will
be derived by a limiting process, starting from the equations for the
discrete case.

Following Hofbauer and Sigmund \cite{HS1}, we consider an infinite
population of players and a set of strategies $S_{1},\cdots ,S_{N}$. When a
player $A$ opts for strategy $S_{i}$ against $B,$ who uses $S_{j}$, the
payoff to $A$ is taken to be $M_{ij},$ and the payoff to $B$ is $M_{ji}.$
Consequently, if $p_{j}(t)$ is the fraction of the whole population that at
time $t$ plays the strategy $S_{j}$, the average payoff to each player using 
$S_{i}$ is equal to 
\begin{equation}
\Pi _{i}(\mathbf{p})\equiv \sum_{j=1}^{N}M_{ij}p_{j}.  \label{indpay}
\end{equation}
The average payoff for the whole population therefore is 
\begin{equation}
\overline{\Pi }(\mathbf{p})=\sum_{i=1}^{N}\Pi _{i}(\mathbf{p}%
)p_{i}=\sum_{i,j=1}^{N}p_{i}M_{ij}p_{j}.  \label{avfit}
\end{equation}
Now in the course of time, the fraction $p_{i}$ changes at a rate which is
proportional to the difference between the payoff to $p_{i}$ and the average
payoff for the whole population. In addition we assume that for each player
of $\ S_{j}$ there is a transition probability per unit time to make a
spontaneous transfer to strategy $S_{i}$ at a rate given by $W_{ij}$. In
this way the \textit{discrete replicator equation with mutation} is derived: 
\begin{equation}
\dot{p}_{i}=(\Pi _{i}(\mathbf{p})-\overline{\Pi }(\mathbf{p}%
))p_{i}+\sum_{j=1}^{N}(W_{ij}p_{j}-W_{ji}p_{i}).  \label{drep}
\end{equation}
It is easy to show that $S(t)\equiv \sum\limits_{i=1}^{N}p_{i}(t)=1$ for all 
$t$ if $S(0)=1.$

In transforming to a continous strategy space, we replace the discrete index 
$i$ by a continuous variable $s\in D\subset R^{n}$ and the variables $%
p_{i}(t)$ by a probability distribution $P(s,t)$. The payoff matrix $M_{ij}$
must now be replaced by a payoff function $M(s,s^{\prime })$, which gives
the payoff to strategy $s$ when playing strategy $s^{\prime }$.

Eq.(\ref{drep}) now takes the form 
\begin{equation}
\frac{\partial P(s,t)}{\partial t}=(\Pi (s,P)-\overline{\Pi }(P))P(s,t)+%
\mathcal{M}(P,s)  \label{crep}
\end{equation}%
in which the mutation term is equal to 
\begin{equation}
\mathcal{M}(P,s)=\int [W(s|s^{\prime })P(s^{\prime },t)-W(s^{\prime
}|s)P(s,t)]\,\mathrm{d}s^{\prime }.  \label{mut}
\end{equation}%
The average payoff for strategy $s$ and the total average payoff are given
by 
\begin{equation}
\Pi (s,P)=\int M(s,s^{\prime })P(s^{\prime },t)\,\mathrm{d}s^{\prime }
\label{Pisp}
\end{equation}%
and 
\begin{equation}
\overline{\Pi }(P)=\int \Pi (s,P)P(s,t)\,\mathrm{d}s.  \label{Pip}
\end{equation}%
Let us now restrict ourselves to mutations in which only small changes in
the strategies occur and apply the method which is used to derive the
Fokker-Planck equation from a master equation \cite{VK}. For simplicity of
presentation, we restrict ourselves to the case of a one-dimensional
strategy space. Define $\xi $ by $s^{\prime }=s-\xi $ and write the mutation
rate as a function of $s$ and $\xi $%
\begin{equation}
W(s^{\prime }|s)=\widetilde{W}(s,s-s^{\prime })=\widetilde{W}(s,\xi )
\label{mut1}
\end{equation}%
and so 
\begin{equation}
W(s|s^{\prime })=\widetilde{W}(s^{\prime },s^{\prime }-s)=\widetilde{W}%
(s-\xi ,-\xi ).  \label{mut2}
\end{equation}%
Assume now that $\widetilde{W}(s,\xi )$ varies slowly in the first variable $%
s$ and that due to the mutations only small variations in the strategies
will occur. Then $\widetilde{W}(s,\xi )$ is only nonvanishing when $\xi $ is
small. In the mutation term (\ref{mut}), which can be written as 
\begin{equation}
\mathcal{M}(P,s)=\int [\widetilde{W}(s-\xi ,-\xi )P(s-\xi ,t)-\widetilde{W}%
(s,\xi )P(s,t)]\,\mathrm{{d}\xi ,}  \label{mut3}
\end{equation}%
we can now expand the dependence on the first variable in powers of $\xi $
and obtain 
\begin{eqnarray}
\mathcal{M}(P,s) &=&\int [\widetilde{W}(s,-\xi )P(s,t)-\xi \frac{\partial }{%
\partial s}\{\widetilde{W}(s,-\xi )P(s,t)\}+  \label{mut4} \\
&&+\frac{1}{2}\xi ^{2}\frac{\partial ^{2}}{\partial s^{2}}\{\widetilde{W}%
(s,-\xi )P(s,t)\}+\cdots -\widetilde{W}(s,\xi )P(s,t)]\,\mathrm{{d}\xi .} 
\notag
\end{eqnarray}%
Because $\int [\widetilde{W}(s,-\xi )-\widetilde{W}(s,\xi )]P(s,t)\mathrm{{d}%
\xi =0}$, the first and last term cancel, so that to second order we are
left with 
\begin{equation}
\mathcal{M}(P,s)=\frac{\partial }{\partial s}\{\alpha _{1}(s)P(s,t)\}+\frac{1%
}{2}\frac{\partial ^{2}}{\partial s^{2}}\{\alpha _{2}(s)P(s,t)\},
\label{mut5}
\end{equation}%
in which 
\begin{equation}
\alpha _{1}(s)=-\int \xi \widetilde{W}(s,\xi )\,\mathrm{d}\xi {\qquad }\text{%
and}{\qquad }\alpha _{2}(s)=\int \xi ^{2}\widetilde{W}(s,\xi )\,\mathrm{d}%
\xi .  \label{alf}
\end{equation}%
We will further simplify the equations by assuming that the average change
in strategy due to mutations is equal to zero, so $\alpha _{1}(s)=0,$ and
that the average of the square of this change is constant, so $\alpha
_{2}(s)=2\sigma .$

The final form of the \textit{continuous replicator equation }(\ref{crep})%
\textit{\ }then becomes 
\begin{equation}
\frac{\partial P(s,t)}{\partial t}=(\Pi (s,P(t))-\overline{\Pi }%
(P(t)))P(s,t)+\sigma \Delta P(s,t),  \label{rep}
\end{equation}%
where we have restored the correct dimensionality of the strategy space by
replacing $\frac{\partial ^{2}}{\partial s^{2}}$ by the \textit{n}%
-dimensional Laplace operator. $P(s,t)$ should satisfy $P(s,t)\geq 0$ and $%
S(t)\equiv \int_{D}P(s,t)\,$d$s,$ should be equal to unity at all times.
This last condition is fulfilled when we choose Neumann, or reflecting, boundary
conditions:

\begin{equation}
\mathbf{n}.\nabla P(s,t)|_{\partial D}=0,  \label{bound}
\end{equation}%
where $\mathbf{n}$ is the normal to the boundary $\partial D$ of the domain $%
D$.

Indeed, integrating (\ref{rep}) over $D$ and using (\ref{bound}) we find 
\begin{equation}
\frac{dS(t)}{dt}=\overline{\Pi }(P)(1-S(t)),  \label{dst}
\end{equation}%
showing that if $S(0)=1$ then $S(t)=1$ for all times. In the case that $%
\sigma =0$, condition (\ref{bound}) is not required to ensure that $S(t)$
remain constant in time.

Equation (\ref{rep}) is a nonlinear reaction-diffusion
equation, where the reaction term $\Pi (s,P(t))-\overline{\Pi }(P(t))$ is
nonlocal. On the function space of twice continuous space-differentiable and
once time-differentiable functions, we can show that the solution of (\ref%
{rep}) exists for all times. This follows from the assumption that $%
M(s,s^{\prime })$ is bounded on $D$, so that $\left\vert \Pi (s,P(t))-%
\overline{\Pi }(P(t))\right\vert \leq \int_{D\times D}\left\vert
M(s,s^{\prime })\right\vert P(s,t)P(s^{\prime },t)\,\mathrm{d}s\,\mathrm{d}%
s^{\prime }+\int_{D}\left\vert M(s,s^{\prime })\right\vert P(s^{\prime },t)\,%
\mathrm{d}s^{\prime }\leq $

$\max \left\vert M(s,s^{\prime })\right\vert \left\{ \int_{D\times
D}P(s,t)P(s^{\prime },t)\,\mathrm{d}s\,\mathrm{d}s^{\prime
}+\int_{D}P(s^{\prime },t)\,\mathrm{d}s^{\prime }\right\} $

$=2\max \left\vert M(s,s^{\prime })\right\vert $. 

Standard comparison theorems for parabolic equations (Pao \cite{PAO}) complete the
proof. Also, by standard positivity results for parabolic equations, it can
be shown that the when the initial distribution $P(s,0)\geq 0,$ then $P(s,t)\geq 0$
for all times $t.$

Numerical simulations suggest that even stronger results hold. In
particular, we suspect that the solution of Eqs.(\ref{rep}), (\ref{bound}) is
uniformly (in space and time) bounded in terms of the sup-norm of the initial distribution. For one-dimensional strategy spaces, we speculate that the solution will always
converge to a stationary solution. 

For $\sigma =0$, Eq.(\ref{rep}) has been studied by Cressman \cite{CR} and
Oechsler \& Riedel \cite{OR}. They show that Eq.(\ref{rep}) has a unique
solution, for all times, on a large space of distributions, containing
amongst others the Dirac-delta distributions.
\newpage
\section{Symmetric 2x2 games}

In symmetric $2\times 2$ games there are two possible strategies, denoted by 
$I$ and $II$. The payoff to player $A$ is given by the payoff matrix

\begin{center}
$M$= 
\begin{tabular}{|l||l|l|}
\hline
A{$\backslash$}B & I & II \\ \hline\hline
I & a & b \\ \hline
II & c & d \\ \hline
\end{tabular}
.
\end{center}

\subsection{Discrete replicator dynamics}

The discrete replicator dynamics associated with this game consists of an
infinite population where a fraction $x_{1}(t)$ plays the pure strategy $I$
and a a fraction $x_{2}(t)$ plays the pure strategy $II$ . The payoffs to
strategies $I$ and $II$ are given by: 
\begin{equation}
\Pi _{I}(x_{1},x_{2})=ax_{1}+bx_{2}\qquad \text{and}\qquad \Pi
_{II}(x_{1},x_{2})=cx_{1}+dx_{2}.
\end{equation}%
The average payoff to the total population is then 
\begin{equation}
\overline{\Pi }(x_{1},x_{2})=\Pi _{I}(x_{1},x_{2})x_{1}+\Pi
_{II}(x_{1},x_{2})x_{2}=ax_{1}^{2}+(b+c)x_{1}x_{2}+dx_{2}^{2}.
\end{equation}%
With a mutation rate matrix of the form $W=\sigma \left( 
\begin{array}{ll}
0 & 1 \\ 
1 & 0%
\end{array}%
\right) $, $\sigma >0$ and using $x_{1}(t)+x_{2}(t)=1$, this leads to the
following equation for $x_{1}(t):$%
\begin{equation}
\dot{x}_{1}=x_{1}(1-x_{1})(B+(A-B)x_{1})+\sigma (1-2x_{1}),  \label{discrrep}
\end{equation}%
where $A=a-c$ and $B=b-d$. The solutions of Eq.(\ref{discrrep}) for $\sigma
=0$ are summarised in figure 1.

\begin{figure}[hbt]
\centering
\includegraphics[width=8cm]{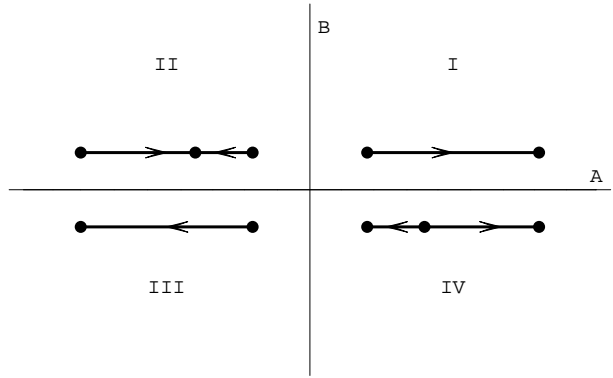}
\caption{The four quadrants of the parameter space}
\end{figure}

A typical example of a game in the quadrant $A>0,B>0$ is the classic
Prisoners Dilemma, where $a=1,b=5,c=0,d=3$. The strategy $I$ corresponds to
'defect' and strategy $II$ to 'cooperate'. Figure 1 shows that for $\sigma
=0 $ the discrete replicator dynamics Eq.(\ref{discrrep}) predicts a final
outcome of $x_{1}=1$, $x_{2}=0$, or 'All Defect'. The effect of the mutation
is to shift the stable solution to a slightly lower value of $x_{1}$.

The Hawk-Dove game (also known as Chicken) has $a=(G-C)/2,b=G,c=0,d=G/2$,
where now strategy $I$ corresponds to 'hawk' (or 'never back down') and $II$
to 'dove' ('allways back down'). When the cost $C$ to the loser of a
hawk-hawk fight is larger than the gain $G$ a hawk makes when confronting a
dove, we have $A<0,B>0$. It follows from Eq.(\ref{discrrep}) that for $%
\sigma =0$ the solution tends to the stable equilibrium $x_{1}=B/(B-A)=G/C$,
describing a population where the strategies co-exist. Also in this case,
the effect of the mutation term is restricted to a small shift in the
location of the stable equilibrium.

The quadrant with $A>0,B<0$ is the domain of the Coordination Games,
exemplified by the situation $a=2,b=0,c=0,d=1.$ In a Coordination Game it is
advantageous for both players to play the same strategy. In a $2\times 2$
game, this leads to two equilibria. Note, however, that in the example
mentioned here the situation where both play strategy $I$ is superior to the
one where both play $II.$ For $\sigma =0$ there are two stable solutions and
one unstable one, and the final outcome depends on the initial situation.
When $x_{1}(0)<\frac{1}{3}$ the solution will ultimately tend to $x_{1}=0$,
otherwise to $x_{1}=1$. In other words, when the initial fraction of
strategy $I$ players is too small, the final population will consist
exclusively of strategy $II$ players, even though this is the less
attractive of the two equilibria.

In the case of Coordination Games, the effect of mutation can qualitatively
change this picture. With the above given values of $a,b,c$ and $d$, we have
plotted the right-hand side of Eq.(\ref{discrrep}) for several values of $%
\sigma $. (figure 2).

\begin{figure}[hbt]
\centering
\includegraphics[width=7cm]{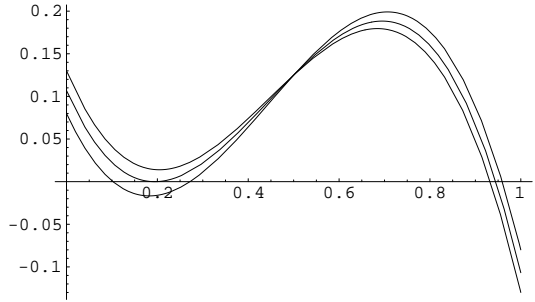}
\caption{The
bifurcation property of Eq.(\protect\ref{discrrep})}
\end{figure}

From this
picture it follows that for values of $\sigma $ larger than a critical value 
$\sigma _{c}$ (in this example $\sigma _{c}\approx 0.11$) the only
equilibrium left is the optimal one near $x_{1}=1$, which is also the
globally attracting solution.

\subsection{Discrete replicator dynamics with a stochastic term}

In \cite{VA}, Eq.(\ref{discrrep}) is studied, where instead of a
deterministic mutation term, a small noise term is added. This leads to the
stochastic equation:

\begin{equation}
dx=G(x)dt+2\sigma \,dW,  \label{discrstoch}
\end{equation}
where $G(x)=x(1-x)(B+(A-B)x)$ and $W(t)$ denotes a Wiener process with zero
mean and unit variance.

In this model, the mutation from one strategy to another does not happen at
a fixed rate, as in the model described by Eq.(\ref{discrrep}), but rather
the fraction $x_{1}(t)$ is changed by a small random amount per time-step.
To describe the outcome of Eq.(\ref{discrstoch}), we consider the evolution
of the probability density $f(x,t).$ The probability that at time $t$ the
fraction of strategy $I$ players lies in the interval $\left[ x,x+\Delta x%
\right] $ is given by $f(x,t)\Delta x$. The equation for $f(x,t)$ is the
Fokker-Planck equation \cite{VK}: 
\begin{equation}
\frac{\partial f(x,t)}{\partial t}=-\frac{\partial }{\partial x}%
(G(x)f(x,t))+\sigma \frac{\partial ^{2}f(x,t)}{\partial x^{2}}.  \label{FP1}
\end{equation}
We assume reflecting boundaries, which yield as boundary conditions: 
\begin{equation}
\frac{\partial f(x=0,t)}{\partial x}=\frac{\partial f(x=1,t)}{\partial x}=0.
\label{bcFP}
\end{equation}
Eq.(\ref{FP1}) with conditions (\ref{bcFP}) has a unique, attracting,
stationary distribution $f^{\ast }(x)$ \cite{VA}.

It is easy to see that this equilibrium distribution is given by 
\begin{equation}
f^{\ast }(x)=C\exp (\frac{1}{\sigma }\int_{0}^{x}G(x^{\prime })\, \mathrm{d}%
x^{\prime }),  \label{fster}
\end{equation}
where the constant $C$ is determined by the condition $\int_{0}^{1}f^{\ast
}(x)dx=1$.

Differentiating $f^{\ast }(x)$ once yields 
\begin{equation}
\frac{df^{\ast }(x)}{dx}=\frac{C}{\sigma }G(x)\exp (\frac{1}{\sigma }%
\int_{0}^{x}G(x^{\prime })\,\mathrm{d}x^{\prime }),
\end{equation}%
from which it follows that the extrema of $f^{\ast }(x)$ are the zeroes of $%
G(x)$, which are the equilibria of the discrete replicator equation (\ref%
{discrrep}) with $\sigma =0$. Differentiating once more gives 
\begin{equation}
\frac{d^{2}f^{\ast }(x)}{dx^{2}}=\frac{C}{\sigma }(G^{\prime }(x)+\frac{%
G^{2}(x)}{\sigma })\exp (\frac{1}{\sigma }\int_{0}^{x}G(x^{\prime })\,%
\mathrm{d}x^{\prime }),
\end{equation}%
so at an equilibrium point $x_{e}$ we have that $\frac{d^{2}f^{\ast
}(x=x_{e})}{dx^{2}}=K\,G^{\prime }(x_{e})$ with $K>0$. Therefore, a stable
equilibrium of Eq.(\ref{discrrep}) with $\sigma =0$ corresponds to a maximum
of $f^{\ast }(x)$, and an unstable equilibrium to a minimum.

In the limit $\sigma \rightarrow 0$, the stationary distribution $f^{\ast
}(x)$ will tend to a point-distribution, where the total probability is
concentrated on one point. In the case of the Hawk-Dove Game and the
Prisoners Dilemma it is clear that this point-distribution is concentrated
on the unique stable equilibrium of the corresponding discrete replicator
equation. For Coordination Games, there are two stable equilibria in the
discrete case. Correspondingly, the stationary distribution has two local
maximuma, namely at $x=0$ and at $x=1$. In \cite{VA} it is proved that for $%
\sigma \rightarrow 0$ one of these two maxima will eventually dominate. The
equilibrium that finally emerges is the one which has the largest basin of
attraction in the discrete case, and is known in the game-theory literature
as the risk-dominant equilibrium.

\subsection{Continuous replicator dynamics}

The two strategy set of $2\times 2$ games can be extended to a continuum of
strategies, each of which indicated by a real number $x\,\epsilon \,[0,1].$
For the payoff function $M(x,x^{\prime })$ we choose a simple interpolation
between the four payoff values of the symmetric $2\times 2$ game: 
\begin{equation}
M(x,x^{\prime })=xx^{\prime }a+x(1-x^{\prime })b+(1-x)x^{\prime
}c+(1-x)(1-x^{\prime })d.  \label{mpay}
\end{equation}

This game can be considered as the underlying discrete $2\times 2$ game
where now mixed strategies are allowed, in the following sense. A strategy $%
x\epsilon \lbrack 0,1]$ means that the player will use pure strategy $I$
with probability $x$. We now assume that two players, using strategies $x$
and $x^{\prime }$ respectively, at one encounter play each other a large
number of times. Eq.(\ref{mpay}) then gives the expectation value of the
payoff to the first player. Because our players will soon become aware of
the Law of Large Numbers, they won't bother with playing against each other,
but at an encounter simply settle for the payoff given by Eq.(\ref{mpay}),
making this a deterministic game.

The expressions for $\Pi (x,P)$ and $\overline{\Pi }(P)$ are easy to
calculate: 
\begin{eqnarray*}
\Pi (x,P) &=&\int_{0}^{1}M(x,x^{\prime })P(x^{\prime },t)\,\mathrm{d}x \\
&=&\int_{0}^{1}(xx^{\prime }a+x(1-x^{\prime })b+(1-x)x^{\prime
}c+(1-x)(1-x^{\prime })d)P(x^{\prime },t)\,\mathrm{d}x^{\prime } \\
&=&d+(b-d)x+(c-d)\overline{x}(t)+(a-b-c+d)x\overline{x}(t)
\end{eqnarray*}%
in which 
\begin{equation}
\overline{x}(t)=\int_{0}^{1}xP(x,t)\,\mathrm{d}x  \label{xav}
\end{equation}%
is the average strategy. Then 
\begin{eqnarray*}
\overline{\Pi }(P) &=&\int_{0}^{1}\Pi (x,P)P(x,t)\,\mathrm{{d}x=} \\
&=&d+(b+c-2d)\overline{x}(t)+(a-b-c+d)\overline{x}^{2}(t),
\end{eqnarray*}%
and 
\begin{eqnarray}
\Pi (x,P)-\overline{\Pi }(P) &=&(b-d)x-(b-d)\overline{x}(t)+(a-b-c+d)(x%
\overline{x}(t)-\overline{x}^{2}(t))  \notag \\
&=&(B+(A-B)\overline{x}(t))(x-\overline{x}(t)),  \label{reac}
\end{eqnarray}%
where $A$ and $B$ are the same parameters as defined in section 3.1.

Eq.(\ref{rep}) for $P(x,t)$ now becomes: 
\begin{equation}
\frac{\partial P(x,t)}{\partial t}=(B+(A-B)\overline{x}(t))(x-\overline{x}%
(t))P(x,t)+\sigma \frac{\partial ^{2}P(x,t)}{\partial x^{2}},  \label{full}
\end{equation}
with boundary conditions: 
\begin{equation}
\left( \frac{\partial P(x,t)}{\partial x}\right) _{x=0}=\left( \frac{%
\partial P(x,t)}{\partial x}\right) _{x=1}=0,  \label{bcpd}
\end{equation}
and an intial distribution $P(x,0)$.

\subsubsection{The equation without mutation}

As was noted before, when considering Eq.(\ref{full}) without mutation, i.e.
with $\sigma =0$, it is not necessary to impose the boundary conditions (\ref%
{bcpd}) in order to ensure that $\int_{0}^{1}P(x,t)\,$d$x$ remain constant.
In \cite{CR} it is shown that 
\begin{equation}
\frac{\partial P(x,t)}{\partial t}=(B+(A-B)\overline{x}(t))(x-\overline{x}%
(t))P(x,t),\;\;\;P(x,0)=P_{0}(x)  \label{geenmut1d}
\end{equation}%
has a unique solution for all $t>0$. In this subsection we will analyse the
asymptotic behaviour of the solution of Eq.(\ref{geenmut1d}) as $%
t\rightarrow \infty $.

Firstly, the equation for the average $\overline{x}(t)$ is given by: 
\begin{equation}
\frac{d\overline{x}}{dt}=(B+(A-B)\overline{x}(t))(\overline{x^{2}}(t)-(%
\overline{x}(t))^{2}),\;\;\;x(0)=\int_{0}^{1}xP_{0}(x)\mathrm{d}x,
\label{xstreep}
\end{equation}
where $\overline{x^{2}}(t)=\int_{0}^{1}x^{2}P(x,t)$d$x$. The factor $%
\overline{x^{2}}(t)-(\overline{x}(t))^{2}$ is always positive. For the four
different regions of the $(A,B)$ parameter plane, as illustrated in figure
1, Eq.(\ref{xstreep}) implies the following.

\begin{description}
\item $(A,B)\in I$: in this case $(B+(A-B)\overline{x}(t))>0$ for all $t$,
so $\overline{x}(t)$ is an increasing function. Because $P(x,t)=0$ for $%
x\notin \lbrack 0,1]$, the distribution will accumulate at $x=1$ and $%
\lim_{t\rightarrow \infty }\overline{x}(t)=1$.

\item $(A,B)\in II$: Eq.(\ref{xstreep}) now has an attractive fixed point at 
$\overline{x}=B/(B-A)$. Therefore, $\lim_{t\rightarrow \infty }\overline{x}%
(t)=B/(B-A)$

\item $(A,B)\in III$: similar to $I$, but now $\lim_{t\rightarrow \infty }%
\overline{x}(t)=0$.

\item $(A,B)\in IV:$Eq.(\ref{xstreep}) has a repelling fixed point at $%
\overline{x}=B/(B-A)$. It follows that $\lim_{t\rightarrow \infty }\overline{%
x}(t)=0$ if $\overline{x}(0)<B/(B-A)$ and $\lim_{t\rightarrow \infty }%
\overline{x}(t)=1$ if $\overline{x}(0)>B/(B-A).$
\end{description}

\bigskip

In the cases $I$, $III$ and $IV$ the limiting distributions $P_{\infty
}(x)=\lim_{t\rightarrow \infty }P(x,t)$ are point-distributions concentrated
on one of the endpoints of the domain $[0,1]$. This is comparable with the
results in the model described in section 3.2 when $\sigma \rightarrow 0$:
eventually all players will play either one or the other of the pure
strategies. The situation is quite different for case $II$, i.e. for
Hawk-Dove type games. We will show that in this case, $P_{\infty }(x)$
depends on the initial condition $P_{0}(x)$, but is in general non-zero on
all of $[0,1]$. This is a dramatic difference with the above mentioned
model. There, the population is divided into a fraction $B/(B-A)$ of the
population who play pure strategy $I$ and the rest who play strategy $II$.
In our model, where the players have access to a continuum of mixed
strategies, we do not find that everybody plays the mixed strategy $%
x=B/(B-A) $ (play strategy $I$ with probability $B/(B-A)$), as might be
expected. Rather, the final outcome is a population who's members play a
broad range of strategies, although the average value of the strategies
played is $\overline{x}=B/(B-A)$. The details are as follows.

\bigskip Every distribution $P(x)$ with $\int_{0}^{1}xP(x)\, \mathrm{d}x=%
\overline{x}=B/(B-A)$ is a solution of Eq.(\ref{geenmut1d}). We will now
show that this set of distributions is attractive.\newline
Let $\lambda (t)=(B+(A-B)\overline{x}(t))$. The solution of Eq.(\ref%
{geenmut1d}) then is 
\begin{equation}
P(x,t)=P_{0}(x)\exp (\int_{0}^{t}\lambda (t^{\prime })(x-\overline{x}%
(t^{\prime }))\,\mathrm{{d}t.}  \label{solgeenmut}
\end{equation}
Using $\int_{0}^{1}P(x,t)$d$x=1$ and writing $g(t)=\int_{0}^{t}\lambda
(t^{\prime })$d$t$, it follows that 
\begin{equation}
\exp (\int_{0}^{t}\lambda (t^{\prime })\overline{x}(t^{\prime
})dt)=\int_{0}^{1}P_{0}(x)\,e^{xg(t)\,\,}\mathrm{{d}x.}  \label{lambda}
\end{equation}
Therefore 
\begin{equation}
P(x,t)=\frac{P_{0}(x)\,e^{xg(t)}}{\int_{0}^{1}P_{0}(x)\,e^{xg(t)}\,\mathrm{{d%
}x}}.  \label{solgeenmut2}
\end{equation}
Differentiating relation (\ref{lambda}) with respect to $t$ yields 
\begin{equation*}
\lambda (t)\,\overline{x}(t)\exp (\int_{0}^{t}\lambda (t^{\prime })\overline{%
x}(t^{\prime })\,\mathrm{d}t)=g^{\prime
}(t)\int_{0}^{1}xP_{0}(x)\,e^{xg(t)}\,\mathrm{d}x.
\end{equation*}
Using $g^{\prime }(t)=\lambda (t)$, and Eq.(\ref{lambda}), this reduces to 
\begin{equation}
\,\overline{x}(t)=\frac{\int_{0}^{1}xP_{0}(x)\,e^{xg(t)}\,\mathrm{d}x}{%
\int_{0}^{1}P_{0}(x)\,e^{xg(t)}\,\mathrm{d}x}.  \label{solxstreep}
\end{equation}
Since $\lim_{t\rightarrow \infty }\overline{x}(t)=B/(B-A)$, then $%
g(t)=\int_{0}^{t}(B+(A-B)\overline{x}(t^{\prime }))\,$d$t^{\prime }$ either
tends to a finite limit or diverges. Using the change of variable $u=x\,g(t)$%
, we can write:

\begin{equation*}
\,\overline{x}(t)=\frac{1}{g(t)}\frac{\int_{0}^{g(t)}uP_{0}(\frac{u}{g(t)}%
)\,e^{u}\,\mathrm{{d}u}}{\int_{0}^{g(t)}P_{0}(\frac{u}{g(t)})\,e^{u}\,%
\mathrm{{d}u}}.
\end{equation*}
Therefore, when $g(t)\rightarrow \infty $, then 
\begin{equation*}
\,\overline{x}(t)=\lim_{\varepsilon \downarrow 0}\frac{\varepsilon
\int_{0}^{1/\varepsilon }uP_{0}(\varepsilon u)\,e^{u}\,\mathrm{d}u}{%
\int_{0}^{1/\varepsilon }P_{0}(\varepsilon u)\,e^{u}\,\mathrm{d}u}%
=\lim_{\varepsilon \downarrow 0}\frac{\varepsilon \int_{0}^{1/\varepsilon
}u\,e^{u}\,\mathrm{d}u}{\int_{0}^{1/\varepsilon }\,e^{u}\,\mathrm{d}u}=1.
\end{equation*}
By a similar argument, when $g(t)\rightarrow -\infty $ then $\overline{x}%
(t)\rightarrow 0$. Since $\lim_{t\rightarrow \infty }\overline{x}(t)=B/(B-A)$
is neither $0$ nor $1$, it follows that $\lim_{t\rightarrow \infty
}g(t)=g_{0}$ is finite. Therefore, as $t\rightarrow \infty $ the solution
Eq.(\ref{solgeenmut2}) tends to 
\begin{equation*}
P_{\infty }(x)=\frac{P_{0}(x)\,e^{xg_{0}}}{\int_{0}^{1}P_{0}(x)\,e^{xg_{0}}\,%
\mathrm{d}x},
\end{equation*}
which is clearly not a point-distribution.

The value of $g_{0}$ can be found from Eq.(\ref{solxstreep}), which in the
limit $t\rightarrow \infty $ reads: 
\begin{equation*}
\frac{B}{B-A}=\frac{\int_{0}^{1}xP_{0}(x)\,e^{xg_{0}}\,\mathrm{d}x}{%
\int_{0}^{1}P_{0}(x)\,e^{xg_{0}}\,\mathrm{d}x}.
\end{equation*}%
This equation for $g_{0}$ can be solved numerically for given values of $A$
and $B$ and a given initial distribution $P_{0}(x)$. In figure 3, two examples,
both with $A=-3$ and $B=7,$ are given for different initial distributions,
shown in the left column. In the right column the final distributions are
plotted.

\begin{figure}[hbt]
\centering
\includegraphics[width=8cm]{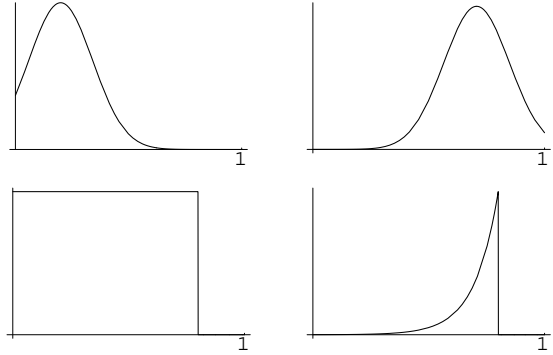}
\caption{Initial and final distributions
without mutations}
\end{figure}

\subsubsection{Stationary solutions of the full equation}

Numerical experiments show that all solutions of the full equation (\ref%
{full}) converge to a time-independent solution. The equation for these
stationary solutions is given by: 
\begin{equation}
\sigma \frac{d^{2}P(x)}{dx^{2}}+(B+(A-B)\overline{x})(x-\overline{x})P(x)=0,
\label{fullstat}
\end{equation}
\begin{equation}
P^{\prime }(0)=P^{\prime }(1)=0,\qquad P(x)\geq 0,\qquad \int_{0}^{1}P(x)\,%
\mathrm{{d}}x=1. \label{statbound}
\end{equation}
Rather than giving a definition of $\overline{x}$ in terms of $P(x)$, we
take $\overline{x}$ to be a free parameter and impose the conditions (\ref%
{statbound}). Integrating Eq. (\ref{fullstat}) over $x$ from $0$ to $1$ then
yields: 
\begin{equation}
(B+(A-B)\overline{x})(\int_{0}^{1}xP(x)\,\mathrm{d}x-\overline{x})=0.
\end{equation}
Assuming that $B+(A-B)\overline{x}\neq 0$, the equality 
\begin{equation}
\overline{x}=\int_{0}^{1}xP(x)\,\mathrm{d}x.
\end{equation}
then follows automatically.

Let $s(\overline{x})=sign(B+(A-B)\overline{x})$ and $\kappa (\overline{x})=|%
\frac{B+(A-B)\overline{x}}{\sigma }|^{1/3}$. Then the solution of Eq. (\ref%
{fullstat}) is given by: 
\begin{equation}
P(x)=a\,Ai[-s(\overline{x})\kappa (\overline{x})(x-\overline{x})]+b\,Bi[-s(%
\overline{x})\kappa (\overline{x})(x-\overline{x})],  \label{solgen}
\end{equation}%
where the Airy functions $Ai(z)$ and $Bi(z)$ are the standard linearly
independent solutions of $y^{\prime \prime }(z)-zy(z)=0$. These functions
are plotted in figure 4.

\begin{figure}[hbt]
\centering
\includegraphics[width=7cm]{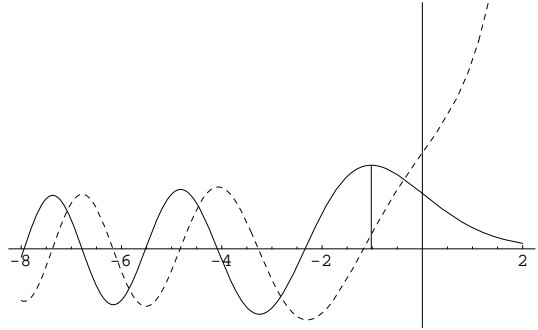}
\caption{Airy
functions $Ai(z)$ and $Bi(z)$}
\end{figure}

The position $%
-\eta _{0}$ of the first maximum of $Ai(z)$ is indicated by a vertical line
segment. The curve for $Bi(z)$ is dashed. Imposing the boundary conditions $%
P^{\prime }(0)=P^{\prime }(1)=0$ gives 
\begin{equation*}
a\,Ai^{\prime }[s(\overline{x})\kappa (\overline{x})\overline{x}%
]+b\,Bi^{\prime }[s(\overline{x})\kappa (\overline{x})\overline{x}]=0
\end{equation*}%
\begin{equation*}
a\,Ai^{\prime }[s(\overline{x})\kappa (\overline{x})(\overline{x}%
-1)]+b\,Bi^{\prime }[s(\overline{x})\kappa (\overline{x})(\overline{x}-1)]=0,
\end{equation*}%
so that a non-trivial solution only exists if $\overline{x}$ is a solution
to the \textquotedblright eigenvalue equation\textquotedblright : 
\begin{equation}
Ai^{\prime }[s(\overline{x})\kappa (\overline{x})\overline{x}]Bi^{\prime }[s(%
\overline{x})\kappa (\overline{x})(\overline{x}-1)]-Bi^{\prime }[s(\overline{%
x})\kappa (\overline{x})\overline{x}]Ai^{\prime }[s(\overline{x})\kappa (%
\overline{x})(\overline{x}-1)]=0.  \label{eigenval}
\end{equation}%
Corresponding to a solution $\overline{x}$ of Eq. (\ref{eigenval}), we find
that the solution (\ref{solgen}) can be written as: 
\begin{gather}
P(x)=  \label{solspec} \\
=c(Bi^{\prime }[s(\overline{x})\kappa (\overline{x})\overline{x}]Ai[s(%
\overline{x})\kappa (\overline{x})(\overline{x}-x)]-Ai^{\prime }[s(\overline{%
x})\kappa (\overline{x})\overline{x}]Bi[s(\overline{x})\kappa (\overline{x})(%
\overline{x}-x)]),  \notag
\end{gather}%
in which $c$ is determined by the normalisation condition. We found that,
although the eigenvalue equation (\ref{eigenval}) can have many solutions,
there will be at most three that correspond to a distribution (\ref{solspec}%
) with the property that $P(x)\geq 0$ for all $x\in \lbrack 0,1]$. For
values of A and B in the regions I and II of figure 1, typified by PD- and
HD-games, there is only one solution. The distributions calculated for $%
\sigma =0.001,$ are shown in figure 5. 

\begin{figure}[hbt]
\centering
\includegraphics[width=12cm]{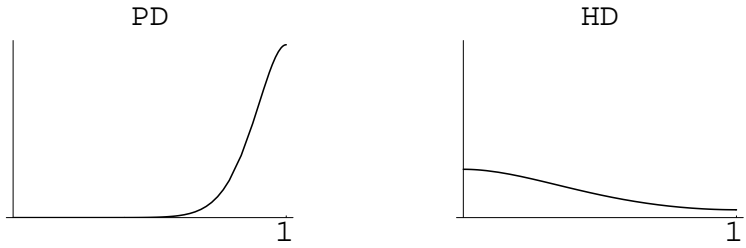}
\caption{Stationary distributions for PD- and HD-games}
\end{figure}

For the PD-case
the parameters are $A=1,$ $B=2$ and $\overline{x}=0.901.$ For the HD-case
they are $A=-2,$ $B=1$ and $\overline{x}=0.342.$ For (A,B)$\,\epsilon \,IV$, 
\textit{i.e.,} for Coordination Games, there exist three solutions for this
value of $\sigma $. The corresponding distributions are plotted in one
picture (figure 6).

\begin{figure}[hbt]
\centering
\includegraphics[width=7cm]{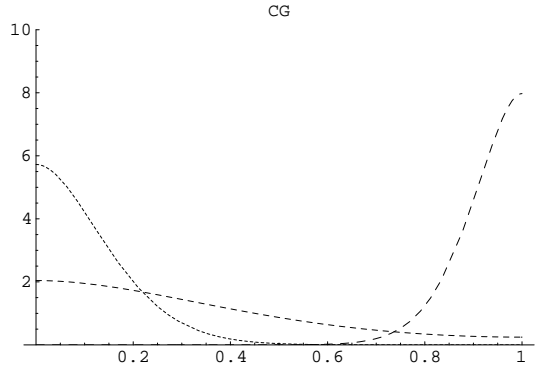}
\caption{Stationary
distributions for CG-game}
\end{figure}

With $A=2$ and $%
B=-1$ we find for the average strategies $\overline{x}=0.118,$ $\overline{x}%
=0.323$ and $\overline{x}=0.915.$

We now study the behaviour of the solution of (\ref{fullstat}) and (\ref%
{statbound}) for $\sigma \rightarrow 0$. First consider the case that $%
|B+(A-B)\overline{x}|>const$, independent of $\sigma $. Then $\kappa (%
\overline{x})=|\frac{B+(A-B)\overline{x}}{\sigma }|^{1/3}\rightarrow \infty $
as $\sigma \rightarrow 0$. From figure 4 it is clear that if $s(\overline{x}%
)=sign(B+(A-B)\overline{x})=1$, then $Bi^{\prime }[s(\overline{x})\kappa (%
\overline{x})\overline{x}]\rightarrow \infty $, $Bi^{\prime }[s(\overline{x}%
)\kappa (\overline{x})(\overline{x}-1)]$ remains bounded and $Ai^{\prime }[s(%
\overline{x})\kappa (\overline{x})\overline{x}]\rightarrow 0$, as $\sigma
\rightarrow 0$. This implies that $\overline{x}$ can only be a solution of
Eq.(\ref{eigenval}) if $Ai^{\prime }[s(\overline{x})\kappa (\overline{x})(%
\overline{x}-1)]\rightarrow 0$. From figure 4, it then follows that $\kappa (%
\overline{x})(\overline{x}-1)\rightarrow -\eta _{0}$, where $\eta
_{0}=1.01879\cdots $ is the smallest positive solution of $Ai^{\prime
}(-\eta _{0})=0,$ so that $\overline{x}\rightarrow 1$ and $\kappa (\overline{%
x})\rightarrow |\frac{A}{\sigma }|^{1/3}$. More precisely, we have: 
\begin{equation*}
\overline{x}=1-|\frac{\sigma }{A}|^{1/3}\,\eta _{0}\qquad as\quad \sigma
\rightarrow 0
\end{equation*}%
This solution is consistent with the assumption $s(\overline{x})=1$ if and
only if $A>0$, as in the Prisoners Dilemma and the Cooperation Game. The
corresponding asymptotic expression for $P(x)$ becomes: 
\begin{equation*}
P(x)=c\,Ai[-\kappa (\overline{x})(x-\overline{x})]=c\,Ai[(\frac{A}{\sigma }%
)^{1/3}(1-x)-\eta _{0}].
\end{equation*}%
Note that for $\sigma \rightarrow 0$ this distribution becomes sharply
peaked at $x=1$, with the width of the peak proportional to $\sigma ^{1/3}$.

By a similar reasoning, it is found that for $B<0$ (as in the Cooperation
Game) there exists a solution such that 
\begin{equation*}
\overline{x}=\,|\frac{\sigma }{B}|^{1/3}\,\eta _{0}\qquad \text{and}\quad
P(x)=c\,Ai[|\frac{B}{\sigma }|^{1/3}x-\eta _{0}]
\end{equation*}%
as $\sigma \rightarrow 0$.

In the previous section we found that for $\sigma =0$ the Hawk-Dove Game and
the Cooperation Game have solutions with $\overline{x}=\frac{B}{B-A}$, which
we call a central solution. This motivates us to look for solutions for
which $\kappa (\overline{x})=|\frac{B+(A-B)\overline{x}}{\sigma }|^{1/3}$
remains bounded as as $\sigma \rightarrow 0$. We therefore assume that 
\begin{equation}
\overline{x}=\frac{B}{B-A}+\frac{\alpha }{B-A}\sigma +...\qquad as\quad
\sigma \rightarrow 0,  \label{asxtreep}
\end{equation}%
where $\alpha \in R$ is as yet unknown. Substituting Eq.(\ref{asxtreep})
into Eq.(\ref{eigenval}) and taking the limit $\sigma \rightarrow 0$ yields: 
\begin{equation}
Ai^{\prime }[\frac{B}{B-A}\alpha ^{1/3}]Bi^{\prime }[\frac{A}{B-A}\alpha
^{1/3}]-Bi^{\prime }[\frac{B}{B-A}\alpha ^{1/3}]Ai^{\prime }[\frac{A}{B-A}%
\alpha ^{1/3}]=0,  \label{alpha}
\end{equation}%
where $\alpha ^{1/3}$ is understood to mean $sign(\alpha )|\alpha |^{1/3}$.
This is the equation from which $\alpha $ must be solved. Although Eq.(\ref%
{alpha}) has many zeroes, only one corresponds to a positive distribution
given by: 
\begin{equation}
P(x)=c(Bi^{\prime }[\beta \alpha ^{1/3}]Ai[\alpha ^{1/3}(x-\beta
)]-Ai^{\prime }[\beta \alpha ^{1/3}]Bi[\alpha ^{1/3}(x-\beta )]),
\label{palpha}
\end{equation}%
with $\beta =\frac{B}{B-A}.$ For $A=-2$ and $B=1,$ (a Hawk-Dove game), it is
plotted in figure 7.

\begin{figure}[hbt]
\centering
\includegraphics[width=6cm]{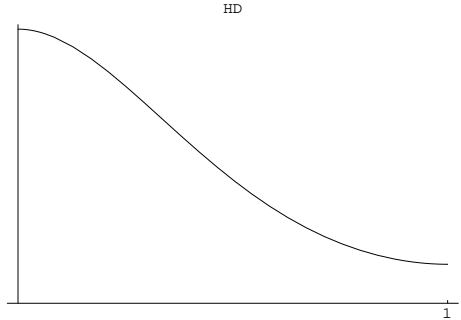}
\caption{Asymptotic
HD-distribution for $\protect\sigma \rightarrow 0$}
\end{figure}

We note that Eqs. (\ref{alpha}) and (\ref{palpha}) are invariant under $%
A\rightarrow -A$, $B\rightarrow -B$. Therefore, as $\sigma \rightarrow 0$,
we find the same central solution for the Coordination Game with $A=2$ and $%
B=-1$. From numerical simulations we find, however, that the central
solution is stable in the Hawk-Dove Game, but unstable in the Coordination
Game.

\bigskip

Away from the limit $\sigma \rightarrow 0$, we can track the fate of the
central solution as $\sigma $ grows. Since the full eigenvalue-equation (\ref%
{eigenval}) is not invariant under $A\rightarrow -A$, $B\rightarrow -B$, the
situation is different for Hawk-Dove Games as opposed to Coordination Games.
For Hawk-Dove Games, we find that the central solution persists for all
values of the diffusion coefficient $\sigma $. However, for Coordination
Games we find that above a critical value $\sigma _{c}$ of $\sigma $, the
unstable central solution disappears, together with one of the two stable
solutions, leaving only one attracting, stationary solution. When $A<|B|$,
the solution at $x=1$ remains, and when $A>|B|$, the solution near $x=0$
survives. This can be summarised by saying that for large enough values of $%
\sigma $, the only attracting solution is a stationary solution near the
risk-dominated solution of the discrete equation. The bifurcation process is
illustrated in figure 8,where for two values of $\sigma $ the left-hand-side of Eq.(\ref{eigenval}) is plotted as a
function of $\overline{x}(\sigma ).$ For $\sigma =0.0035$ this function has
three zeroes, whereas for $\sigma =0.0039$ there is only one$.$ This closely
resembles the situation in the discrete case as described in section 2.
However, the critical values for $\sigma $ in the discrete case and in the
continuous case are not comparable ($0.11$ vs. $0.0037$).

\begin{figure}[hbt]
\centering
\includegraphics[width=8cm]{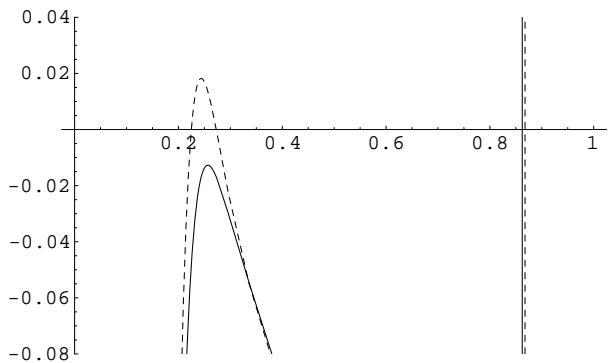}
\caption{Bifurcation of CG-games}
\end{figure}

\subsubsection{Summary}

Here we summarise and compare the results of the three types of games, using
the discrete model, the discrete model with a stochastic term and the
continuous model with deterministic mutation.

The Prisoners Dilemma is straightforward. In the discrete case the
population will eventually play All Defect. This does not change when the
strategy-space is made continuous and adding deterministic mutation simply
changes the limiting delta-distribution to a finite peak with width
proportional to $\sigma ^{1/3}$. The contrast with the corresponding
stochastic equation is that there the width of the peak is narrower and
proportional to $\sigma ^{1/2}$.

In the discrete version of the Coordination Game, the population will
eventually play either $x_{1}=0$ or $x_{1}=1$, where the final outcome
depends on the intial condition. If the mutation rate is larger than a
certain threshhold, only the stable solution around the risk-dominant
solution remains. This result remains the same in the continuous case.
Depending on the initial distribution of strategies, the final distribution
will be sharply peaked (width of peak proportional to $\sigma ^{1/3}$)
around either $x=0$ or $x=1$. When the mutation rate becomes larger than a
critical value, only the distribution around the risk-dominant solution
remains. We note that this critical value is much smaller in the continuous
case than in the discrete case. The stochastic equation has only one,
attracting, stationary solution. When the mutation rate $\sigma $ is small,
this distribution is almost completely concentrated around the risk dominant
solution, where again the width of the distribution is proportional to $%
\sigma ^{1/2}$.

The Hawk-Dove Game shows the following behaviour. In the discrete case,
there is an asymptotically stable solution for all values of $\sigma >0$,
which for $\sigma =0$ has the value $x_{1}=B/(B-A)$. The stochastic equation
has a unique attracting stationary solution, peaked around $x=B/(B-A)$ and
with a width proportional to $\sigma ^{1/2}$. Without mutation, $\sigma =0$,
the continuous equation has an asymptotically stable invariant set of
solutions, consisting of all distributions $P(x)$ with average $\overline{x}%
=B/(B-A)$. Depending on the initial distribution, the solution converges to
a member of this set. When the mutation term is added, $\sigma >0$, only
one, attracting, solution remains. In the limit $\sigma \rightarrow 0$, this
solution converges to (\ref{palpha}). This solution is not concentrated on a
single point, but has the whole interval $[0,1]$ as support. Such a solution
is sometimes referred to as polymorphic. For small $\sigma $, numerical
experiments show that a starting distribution initially converges to a
distribution close to the solution it would reach if $\sigma =0$, but then
slowly (on a time-scale of $1/\sigma $) evolves to the unique limiting
solution. This behaviour resembles that of a singulary perturbed ordinary
differential equation, which in the unperturbed case has an attracting set
of fixed points, and where after adding the perturbation, the attracting
invariant set survives. In this invariant set we would have, in this
analogy, one attracting fixed point left.

\section{The Ultimatum Game}

This game has attracted a great deal of interest, mainly as a model to
explain the occurrence of strong reciprocity in populations of supposedly
selfish individuals.(Binmore \cite{BI}, Fehr \& Gachter \cite{FG}). Here
"strong reciprocity" means the willingness to share, but also to punish
egotistical behaviour in others, even at a cost to oneself. See Bowles \&
Gintis \cite{BG}.

The game is played by two players. The first player is given a certain
amount of money and proposes a split of this money with the second player.
This second player has the choice between accepting the offer, or rejecting
it, in which case neither of the two players will receive anything. An
obvious strategy for the second player is to accept every offer, since
something is better than nothing. Realising this, the first player will
maximise his share by offering the lowest possible amount to the second
player. This (combined) strategy is sometimes referred to as
\textquotedblright the rational solution\textquotedblright\ (Page \& Nowak 
\cite{PN}), \textquotedblright the subgame-perfect
equilibrium\textquotedblright\ (Seymour \cite{SE}) or the strategy of \
\textquotedblright Homo Economicus\textquotedblright\ (Bowles \& Gintis \cite%
{BO}).

An evolutionary version of this game, taking into account mutations, was
studied in \cite{PN} and by Nowak et al. \cite{NPS}, using adaptive
dynamics. In adaptive dynamics models, the population is assumed to always
be monomorphic, \textit{i.e.}, everybody plays the same strategy. Every now
and then, a mutant is introduced. If the strategy of the mutant is more
succesful than the resident strategy, it will quickly spread in the
population, thus becoming the new resident strategy. For the Ultimatum Game
it was found that in the absence of any restrictions, the solution of the
adaptive dynamics model indeed converged to the \textquotedblright rational
solution\textquotedblright .

The Ultimatum Game was also studied by Seymour \cite{SE}, using replicator
dynamics. He included mutations as a given, exogeneous term and found that
other solutions can emerge, far from the ''subgame-perfect solution'',
depending on the form and intensity of the mutation function.

\bigskip

We model the Ultimatum Game as follows. The strategy space is $S=[0,1]\times
\lbrack 0,1]$, where a stategy $s=(x,y)\in S$ means that the player, in the
role of nr. 1, will offer a fraction $x$, while in the role of nr. 2 he will
reject any offer lower than $y$.

In one round, the players will play the role of nr. 1 and nr. 2 alternately.
This leads to a payoff function giving the payoff to strategy $s$ when
playing strategy $s^{\prime }$ (the factor $1/2$ has been omitted): 
\begin{equation}
M(s,s^{\prime })=M(x,y|x^{\prime },y^{\prime })=(1-x)\Theta (x-y^{\prime
})+x^{\prime }\Theta (x^{\prime }-y),  \label{mug}
\end{equation}%
in which the Heaviside function is defined by 
\begin{equation}
\Theta (z)=\left\{ 
\begin{array}{c}
1\quad \text{if}\mathrm{\quad }z \geq 0 \\ 
0\quad \text{if}\mathrm{\quad }z<0%
\end{array}%
\right.  \label{th}
\end{equation}

Before writing down the full replicator equation (\ref{rep}) for this case
we first introduce a number of abbreviations: 
\begin{equation}
H(x,t)=\int_{0}^{1}P(x,y,t)\,\mathrm{d}y\quad \text{and}\quad
V(y,t)=\int_{0}^{1}P(x,y,t)\,\mathrm{d}x,  \label{HV}
\end{equation}%
which are normalised as 
\begin{equation}
\int_{0}^{1}H(x,t)\,\mathrm{d}x=\int_{0}^{1}V(y,t)\,\mathrm{d}y=1.
\label{HVN}
\end{equation}%
$Q(x,t)$ and $R(y,t)$ are defined by 
\begin{equation}
Q(x,t)=\int_{0}^{x}V(y,t)\,\mathrm{d}y\quad \text{and}\quad
R(y,t)=\int_{y}^{1}xH(x,t)\,\mathrm{d}x,  \label{QR}
\end{equation}%
so that 
\begin{equation}
Q(0,t)=0\quad \text{and}\quad Q(1,t)=1\quad \text{and}\quad R(1,t)=0.
\label{QR1}
\end{equation}%
In terms of these functions the local and global averages take the form 
\begin{equation}
\Pi (x,y,P)=(1-x)Q(x,t)+R(y,t)  \label{UGp}
\end{equation}%
and 
\begin{equation}
\overline{\Pi }(P)=c_{1}(t)+c_{2}(t),  \label{pcc}
\end{equation}%
with 
\begin{equation}
c_{1}(t)=\int_{0}^{1}(1-x)Q(x,t)H(x,t)\,\mathrm{d}x\quad \quad
c_{2}(t)=\int_{0}^{1}R(y,t)V(y,t)\,\mathrm{d}y.  \label{cc}
\end{equation}%
At last the replicator equation becomes 
\begin{eqnarray}
\frac{\partial P(x,y,t)}{\partial t}
&=&[(1-x)Q(x,t)+R(y,t)-c_{1}(t)-c_{2}(t)]P(x,y,t)+  \label{UGrep} \\
&&+\sigma \Delta P(x,y,t),  \notag
\end{eqnarray}%
in which $\Delta $ is the two-dimensional Laplace operator. The boundary
condition is 
\begin{equation}
\nabla P\cdot \mathbf{n}=0\quad \text{on\ the\ boundary.}  \label{bnd}
\end{equation}%
In what follows we will restrict ourselves to solutions which can be written
as the product of two normalised functions of $(x,t)$ and of $(y,t)$
respectively. It \ then necessarily follows that 
\begin{equation}
P(x,y,t)=H(x,t)V(y,t).  \label{phv}
\end{equation}%
With this restriction we easily show, by integrating Eq.(\ref{UGrep}) over $%
y $, that 
\begin{equation}
\frac{\partial H(x,t)}{\partial t}=[(1-x)Q(x,t)-c_{1}(t)]H(x,t)+\sigma \frac{%
\partial ^{2}H(x,t)}{\partial x^{2}}.  \label{Hrep}
\end{equation}%
The boundary conditions are 
\begin{equation}
\frac{\partial H(x=0,t)}{\partial x}=\frac{\partial H(x=1,t)}{\partial x}%
=0\quad \text{for\ all}\mathrm{\;}t.  \label{Hbnd}
\end{equation}%
Integration over $x$ of Eq.(\ref{UGrep}) leads to 
\begin{equation}
\frac{\partial V(y,t)}{\partial t}=[R(y,t)-c_{2}(t)]V(y,t)+\sigma \frac{%
\partial ^{2}V(y,t)}{\partial y^{2}},  \label{Vrep}
\end{equation}%
with boundary conditions 
\begin{equation}
\frac{\partial V(y=0,t)}{\partial y}=\frac{\partial V(y=1,t)}{\partial y}%
=0\quad \text{for\ all}\mathrm{\;}t.  \label{Vbnd}
\end{equation}%
We note that for $\sigma =0$ these equations are the same as those studied
in \cite{SE}, where it is assumed that there are two separate populations of
players, one where all members always play the role of nr. 1 and the other
with nr. 2 players.

\subsection{The equation without mutation}

For $\sigma =0$ the equations become: 
\begin{eqnarray}
\frac{\partial H(x,t)}{\partial t} &=&[(1-x)Q(x,t)-c_{1}(t)]H(x,t)  \notag \\
\frac{\partial V(y,t)}{\partial t} &=&[R(y,t)-c_{2}(t)]V(y,t).  \label{nomut}
\end{eqnarray}%
Similar to the Hawk-Dove Game, we identify a set of stationary solutions: 
\begin{eqnarray}
H_{0}(x) &=&\delta (x-\overline{x})  \notag \\
V_{0}(y) &=&\left\{ 
\begin{array}{c}
v(y)\quad \text{if}\quad y<\overline{x} \\ 
0\quad \quad \text{if}\quad y>\overline{x}%
\end{array}%
\right.  \notag \\
c_{1} &=&1-\overline{x}\quad ,\quad c_{2}=\overline{x}  \label{stat}
\end{eqnarray}%
where $v(y)$ is an arbitrary function with $\int_{0}^{\overline{x}}v(y)\,$d$%
y=1$ and $\delta (z)$ is the Dirac-$\delta $ distribution. The solution (\ref{stat}) is easily checked, by
noting that $R(y)=\overline{x}\Theta (\overline{x}-y)$ and that $Q(\overline{%
x})=1$. We have also used $%
z\,\delta (z)\equiv 0$. The interpretation of this solution is clear: player nr. 1 always
offers $\overline{x}$, so player nr. 2 will always receive this amount, as
long as his acceptence threshhold is below $\overline{x}$. The average payoff
is therefore $\overline{\Pi }=c_{1}+c_{2}=1$ and any distribution of the $y$%
-values below $\overline{x}$ is stationary, given this distribution of $x$.
The limit $\overline{x}\rightarrow 0$ corresponds to the subgame-perfect
solution.

To show that (\ref{stat}) represents an attracting set of solutions, we use
the same reasoning as in section 3.3.1, and find 
\begin{equation}
\exp [\int_{0}^{t}c_{1}(t^{\prime })\,\mathrm{d}t^{\prime
}]=\int_{0}^{1}H_{0}(x)\,\exp [(1-x)\int_{0}^{t}Q(x,t^{\prime })\,\mathrm{d}%
t^{\prime }]\,\mathrm{d}x\equiv h(t)  \label{expc1}
\end{equation}%
\begin{equation}
H(x,t)=\frac{H_{0}(x)}{h\left( t\right) }\exp
[(1-x)\int_{0}^{t}Q(x,t^{\prime })\,\mathrm{d}t^{\prime }]  \label{Huitdr}
\end{equation}%
\begin{equation}
\mathrm{\exp }[\int_{0}^{t}c_{2}(t^{\prime })\,\mathrm{d}t^{\prime
}]=\int_{0}^{1}V_{0}(y)\,\exp [\int_{0}^{t}R(x,t^{\prime })\,\mathrm{d}%
t^{\prime }]\,\mathrm{d}y\equiv v(t)  \label{expc2}
\end{equation}%
\begin{equation}
V(y,t)=\frac{V_{0}(y)}{v(t)}\exp [\int_{0}^{t}R(y,t^{\prime })\,\mathrm{d}%
t^{\prime }].  \label{Vuitdr}
\end{equation}%
By differentiating Eq. (\ref{expc1}) and (\ref{expc2}), respectively, we
obtain: 
\begin{equation}
c_{1}(t)=\frac{1}{h(t)}\int_{0}^{1}(1-x)Q(x,t)H_{0}(x)\,\exp
[(1-x)\int_{0}^{t}Q(x,t^{\prime })\,\mathrm{d}t^{\prime }]\,\mathrm{d}x
\label{c1}
\end{equation}%
\begin{equation}
c_{2}(t)=\frac{1}{v(t)}\int_{0}^{1}R(y,t)V_{0}(y)\,\exp
[\int_{0}^{t}R(y,t^{\prime })\,\mathrm{d}t^{\prime }]\,\mathrm{d}y.
\label{c2}
\end{equation}%
Assuming that $V(y,t)$ converges to a stationary distribution (as all
numerical results show), then $(1-x)\int_{0}^{t}Q(x,t^{\prime })\,\mathrm{d}%
t^{\prime }$ converges to a function with a finite number of isolated local
maxima. One of these, say $x=\overline{x}$, is the absolute maximum, and it
follows from Eq.(\ref{Huitdr}) that $H(x,t)$ converges to $\delta (x-%
\overline{x})$. From Eq.(\ref{QR}) it follows that $R(y,t)$ converges to $%
R(y)=\overline{x}\,\Theta (\overline{x}-y)$ and $V(y,t)$ converges to $%
c\,V_{0}(y)\Theta (\overline{x}-y)$, with $c$ a normalization constant. From
Eq.(\ref{cc}) it follows that $c_{1}(t)$ converges to $1-\overline{x}$ and $%
c_{2}(t)$ to $\overline{x}$.\newline
The above considerations show that if the solution of Eq.(\ref{nomut})
converges to a stationary solution, it must be a member of the invariant set
(\ref{stat}). However, the value of $\overline{x}$ cannot be predicted from
the above formula's. Numerical solution of Eq.(\ref{nomut}) shows that for
random initial distributions of $H(x,0)$ and $V(y,0)$ on the whole interval $%
[0,1],$ the functions $H(x,t)$ and $V(y,t)$ indeed approach the form of Eq.(%
\ref{stat}) for $t\rightarrow \infty $. The average strategy $\overline{x},$
based on $100$ simulations, takes values between $0.12$ and $0.30,$ with a
mean value equal to $0.22$ and a standard deviation of $0.04.$ We note that
a uniform distribution of both $H(x,0)$ and $V(y,0)$ also leads to a value
of $\overline{x}=0.22.$

\subsection{The equation with mutation}

We have found numerically that when $\sigma >0$, all solutions of the full equations (\ref{Hrep}) and (\ref{Vrep}) tend
to a unique solution of the stationary equations: 
\begin{equation}
\sigma \frac{{d}^{2}H(x)}{{\ d}x^{2}}+[(1-x)Q(x)-c_{1}]H(x)=0  \label{Hh}
\end{equation}%
and 
\begin{equation}
\sigma \frac{{\ d}^{2}V(y)}{{\ d}y^{2}}+[R(y)-c_{2}]V(y)=0.  \label{Vh}
\end{equation}%
The boundary values are those of Eqs.(\ref{Hbnd}) and(\ref{Vbnd}).\newline
Unfortunately, we have not been able to find closed form expressions for the
solutions of these equations. There are two ways to approximate the stationary solution, which lead to the same
result. First, we numerically solved the full equations(\ref{Hrep}) and (\ref{Vrep}), by
discretising space, solving the resulting coupled set of ordinary
differential equations and considering the solution as $t \rightarrow \infty$. In the second method, we define the following seven functions 
\begin{eqnarray}
F_{1}(x) &=&H(x),\quad F_{2}(x)=\frac{dH(x)}{dx},\quad F_{3}(x)=V(x),\quad
F_{4}(x)=\frac{dV(x)}{dx},  \notag \\
F_{5}(x) &=&Q(x),\quad F_{6}(x)=R(x),\quad F_{7}(x)=\int_{x}^{1}H(x^{\prime
})\,\mathrm{{d}x^{\prime }.}  \label{f7}
\end{eqnarray}%
In terms of these functions and with $k=1/\sigma $ ,the stationary equations
can now be written as 
\begin{eqnarray}
\frac{dF_{1}(x)}{dx} &=&F_{2}(x)  \notag \\
\frac{dF_{2}(x)}{dx} &=&-k[(1-x)F_{5}(x)-c_{1}]F_{1}(x)  \notag \\
\frac{dF_{3}(x)}{dx} &=&F_{4}(x)  \notag \\
\frac{dF_{4}(x)}{dx} &=&-k[F_{6}(x)-c_{2}]F_{3}(x)  \label{feq7} \\
\frac{dF_{5}(x)}{dx} &=&F_{3}(x)  \notag \\
\frac{dF_{6}(x)}{dx} &=&-xF_{1}(x)  \notag \\
\frac{dF_{7}(x)}{dx} &=&-F_{1}(x)  \notag
\end{eqnarray}%
These equation can be solved numerically by starting the integration from
the following values at $x=1$%
\begin{eqnarray}
F_{1}(1) &=&a,\quad F_{2}(1)=0,\quad F_{3}(1)=b,\quad F_{4}(1)=0,  \notag \\
F_{5}(1) &=&1,\quad F_{6}(1)=0,\quad F_{7}(1)=0,  \label{fbnd}
\end{eqnarray}%
and using a standard routine to arrive at the values of these functions in $%
x=0.$ The numbers ($a$,$b$,$c_{1}$,$c_{2})$ are as yet unknown. They should
be chosen in such a way that the boundary conditions at $x=0$ be satisfied, 
\textit{i.e.,} 
\begin{equation}
(F_{2}(0),F_{4}(0),F_{5}(0),F_{7}(0))=(0,0,0,1).  \label{fcon}
\end{equation}%
This matching of four numbers by varying four other numbers should be
possible in many ways. It turns out, however, that the requirement of
positivity of $F_{1}(x)$ and $F_{3}(x)$ in the whole interval $[0,1]$ makes
the solution unique. A root finding routine of Mathematica does the job. For 
$\sigma =0.001$ the stationary solutions $H(x)$ and $V(y)$ are shown in
figure 9. 

\begin{figure}[hbt]
\centering
\includegraphics[width=12cm]{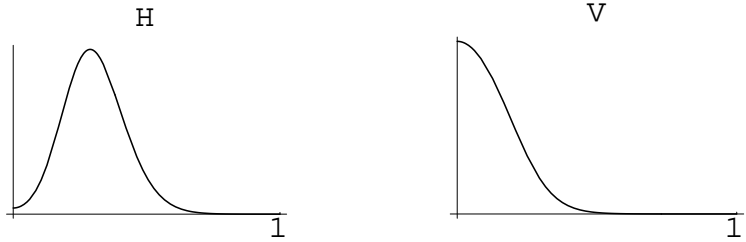}
\caption{Stationary solution
of Eqs.(\protect\ref{Hh}) and (\protect\ref{Vh}) for $\protect\sigma =0.001$}
\end{figure}

We note that $%
H(x)$ has a Gauss-like distribution around a mean value $\overline{x}=0.3172$%
, while $V(y)$ is approximated by the right half of a Gaussian, with its maximum
at $y=0$. For smaller values of $\sigma $, the value of $%
\overline{x}(\sigma )$ and the width of the peak of the $H(x)$-distribution
decrease, but the shape of the distributions is otherwise unchanged.

For values of \ $\sigma $ down to $10^{-9}$ we have calculated the $\sigma $%
-dependence of $\overline{x}.$ In figure 10

\begin{figure}[hbt]
\centering
\includegraphics[width=6cm]{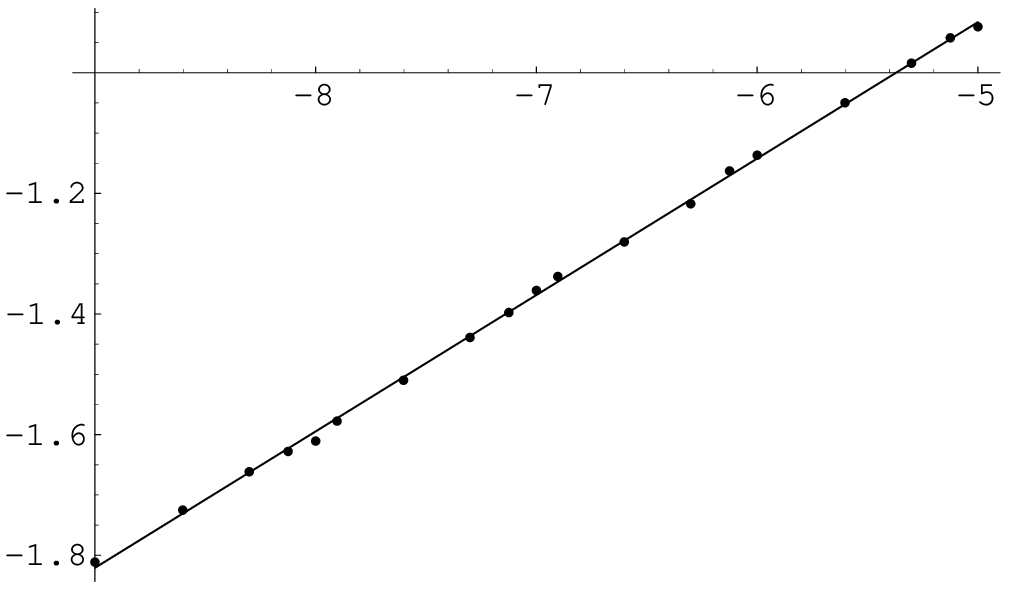}
\caption{Log-log plot of $\overline{x}$ as function of $\protect\sigma $}
\end{figure}

a log-log plot
of this dependence is shown. A good fit of the data points is given by $%
\overline{x}(\sigma )\approx 1.64\times \,\,\sigma ^{0.23}$.

The numerical solution of the full time dependent equations reveals a
dynamical pattern similar to the Hawk-Dove Game of section 2. Initially a
distribution approaches the attracting set (\ref{stat}), after which it
slowly converges to the unique stationary solution, on a time-scale of $%
1/\sigma $ .

\section{Conclusions}

In this paper we have generalised the replicator dynamics of games with
deterministic mutations, as described in \cite{HS1}, to the situation where
players have access to a continuous strategy space. The resulting equation (%
\ref{rep}) has a well defined solution, which, however, is not easy to
analyse in general.

Our first example, the continuous version of $2\times 2$ symmetric games,
already illustrates a number of interesting and perhaps unexpected features
of this equation. Although there is no a priori reason to believe that the
continuous and the discrete strategy version of the same $2\times 2$ game
have anything to do with each other, the similarities between some results
warrant our surprise at the differences in others. The continuous Prisoners
Dilemma and the Coordination Game behave similar to their discrete case
counterparts: the final state is a monomorphic distribution, where every
player in the population plays one of the two pure strategies. The extension
to a continuous strategy space and the inclusion of mutation (which has the
form of a diffusion term) only leads to the existence of small variations
around the single peak of the final distribution. Also in the continuous
Coordination Game we encounter, as in the discrete case, a threshold value
for the mutation term separating a regime with two attracting solutions from
one where only a single attractor exists.

The difference occurs in the Hawk-Dove game. In the discrete version there
is a stable equilibrium with Hawks and Doves coexisting. In the continuous
version this does not translate into a monomorphic distribution around the
mixed strategy corresponding to this equilibrium. Rather, in the unique
limiting distribution the whole range of mixed strategies from pure Hawk to
pure Dove is represented. The attraction to this stationary solution occurs
on two time scales. On a fast time scale, the solution is attracted to the
set of distributions with average corresponding to the equilibrium mixed
strategy of the discrete case. Then, on a slow time scale proportional to
the inverse of the mutation rate, the solution converges to the unique
attractor. This two timescales phenomenon was observed in numerical
simulations, and is currently awaiting a more thorough analysis. Also, there
are many interesting games with three or more strategies (for instance Rock,
Scissors, Paper) of which the continuous version can hold more surprises.%
\newline
\newline
The results of the second example, the Ultimatum Game, are of great interest
to the debate around strong reciprocity and how it could have evolved. Our
model shows that replicator dynamics and a small mutation term can lead to a
final outcome far from the subgame-perfect solution. Take, for instance, a
mutation rate of $10^{-3}.$  So in every time-interval, all players vary their strategies
according to the cold rules of self-interest, after which a small fraction of $%
0.1\%$ of the population, change their strategy just a little bit.
Then we find that an initial population consisting almost entirely of
cynical misers (accept everything and offer nothing), eventually turns into
a world where the average offer is more than $30\%$!

This surprising result can be explained in the following way. Consider a
situation where all proposers offer only a small share to their opponent and these
opponents all have an acceptance threshhold lower than this offer. Now, due to
mutation, some acceptors will demand a share that is slightly larger than what
is being offered. Normally, this would be a suicidal strategy. However, also due
to mutation, there will be amongst the proposers a small set who are willing to
offer slightly more than their colleagues. 
On the one hand, these fairer-minded proposers earn slightly less from the bulk
of the acceptors, but on the other hand they are the only ones to profit from the small group of
high-minded mutants on the acceptor-side. The net result can be that the second
effect dominates and that there will be a tendency towards higher offers.

For sufficiently small mutation rates the dynamics of the Ultimatum Game show the same structure
with two timescales as the continuous Hawk-Dove game. An initial
distribution is quickly attracted to a distribution where the offers are
sharply peaked, and then slowly converges to the unique stationary solution.

In this case too, a more rigorous mathematical analysis is required for a
better understanding of the model. In particular it would be nice to be able
to calculate the value of the exponent in the formula relating the mutation
rate and the average value of the offers, which in this paper we derived
from numerical simulations. For this purpose singular perturbation theory
seems to be an appropriate tool. Furthermore, we have only considered
mutation rates that are the same for the proposers as for the acceptors.
Differentiating between these may also lead to a fuller understanding.

\end{document}